\documentstyle[amssymb,12pt,thmsa,a4,sw20lart]{article}

\input tcilatex
\begin{document}

\topmargin 0pt \oddsidemargin 5mm

\setcounter{page}{1}

\begin{quotation}
\hspace{8cm}{} \vspace{2cm}
\end{quotation}

\begin{center}
{\bf EXCITATION OF NONLINEAR ONE-DIMENSIONAL WAKE WAVES IN UNDERDENSE AND
OVERDENSE MAGNETIZED PLASMA BY A RELATIVISTIC ELECTRON BUNCH}\\ \vspace {5mm}

H. B. Nersisyan$^1$ and S. S. Elbakian$^2$\\ \vspace{1cm} $^1${\em Institute
of Radiophysics and Electronics, Ashtarak-2, 378410, Armenia}

$^2${\em Yerevan Physics Institute, Alikhanian Brothers St. 2, Yerevan
375036, Republic of Armenia}\\ E-mail: hrachya@irphe.am
\end{center}

\vspace {5mm} \centerline{{\bf{Abstract}}}

\begin{quotation}
The excitation of wake waves by a relativistic homogeneous electron bunch
passing through cold magnetized plasma at equilibrium is studied for
arbitrary values of the ratio of bunch density to plasma density. The
analysis is based on the assumption that the magnetic field is sufficiently
strong ($\omega _B\gg \omega _p$, where $\omega _p$ and $\omega _B$ are the
electron plasma and cyclotron frequences respectively). The periodic and
nonperiodic solutions for the momentum of plasma electrons inside and
outside the bunch are analyzed. It was shown that the presence of strong
external magnetic field may increase the amplitude of wake waves and the
maximum transformer ratio, the latter being reached at densities of bunch
lower than that in the absence of magnetic field. The optimum conditions for
obtaining the maximum values of the wake field amplitude and of the
transformer ratio are found.

\newpage 
\end{quotation}

\section{INTRODUCTION}

The plasma wake-field accelerator (PWFA), proposed by Chen, Dawson, Huff,
and Katsouleas$^1$ in 1985, and discussed in a slightly different context by
Fainberg$^2$ as early as 1956, uses electrostatic fields in a plasma wave
driven by a relativistic electron beam. Charged particles can then be
accelerated in this electrostatic wake to ultrahigh energies. The first
experimental observation, at the Argonne National Laboratory Advanced
Accelerator Test Facility, of acceleration by the PWFA mechanism has been
reported in Ref.$^3$ The one-dimensional theory of the nonlinear regime of
the PWFA developed by Rosenzweig$^{4-6}$, Ruth et al.$^7$, Amatuni et al.$%
^{8,9}$ and Khachatryan$^{10}$ predict certain advantages over the linear
regime (for more details see reviews$^{11,12}$). It has been shown that the
transformer ratio, the ratio of the maximum decelerating field inside the
driving electron bunch to the maximum accelerating field found in the wake
of the driver, is enhanced by driving the plasma waves with an
ultrarelativistic electron bunch of density one-half that of the plasma
electrons. This interesting method of obtaining large transformer ratios is
asserted to be more straightforward than alternatives in the linear regime,
and also to lessen the multiple scattering that the accelerating particles
undergo in a PWFA, as the plasma used need not be as dense as the linear
regime requires. The simplicity of the nonlinear scheme for obtaining high
transformer ratios, as well as the inherently high accelerating gradients,
also makes this method interesting as a candidate for high-energy
cosmic-ray-acceleration mechanism.

The high transformer ratios obtained in this nonlinear scheme depend on
driving the plasma electron density waves to extremely large amplitudes. In
contrast to the linear waves, the maximum energy the electrons reach during
the oscillation is many times their rest mass energy. The electron
oscillation period is an increasing function of this maximum energy. As the
amplitude of the wave grows, the wave steepens dramatically, and the
positive excursion in density becomes very large in amplitude and narrow in
time.

An important result of the nonlinear theory is the proof that the wave
breaking limit is $E_{\max }=(mu_b\omega _p/e)\left[ 2(\gamma _b-1)\right]
^{1/2}$ and it is reached at $n_b/n_0\lesssim 1/(2+\gamma _b^{-1})$, where $%
\omega _p=\left( 4\pi n_0e^2/m\right) ^{1/2}$ is the plasma frequency of
electrons, $n_b$ and $n_0$ are the densities of the bunch and plasma
electrons, $\gamma _b=\left( 1-\beta _b^2\right) ^{-1/2}$ is the
relativistic factor of the bunch, $\beta _b=u_b/c$, $u_b$ is the velocity of
the bunch. The Dawson$^{13}$ wave breaking limit is equal to $E_{\max
}\thickapprox 2mu\omega _p/e$ when $\gamma _b\thickapprox 1$ $(\beta _b\ll
1) $.

The one-dimensional relativistic strong waves can be excited in the plasma
by wide relativistic bunches of charged particles or intense laser pulses$%
^{11,14}$ (when $k_pa_0\gg 1$, where $k_p=\omega _p/u_b$, $a_0$ are the
characteristic transverse sizes of bunches or pulses).

The excitation of nonlinear one-dimensional wake waves in underdense and
overdense plasma by a relativistic electron bunch was studied in Ref.$^9$

In the present work an analogous problem has been treated in the presence of
strong external magnetic field ${\bf B}_0$. A qualitatively novel effect of
dependence of the amplitude of wake wave excited by an one-dimensional bunch
on the angle $\alpha $ of magnetic field orientation with respect to the
direction of bunch motion was found. When $\alpha $ is increased from zero
to $\Delta =\arcsin (1$/$\gamma _b)$, then wake waves with larger amplitudes
as well as larger transformer ratios are obtained. Here, the maximum value
of the transformer ratio is attained at lower densities of the bunch in
comparison with the case of ${\bf B}_0=0$.

\section{BASIC\ EQUATIONS}

Our treatment will start with equations of cold relativistic nonlinear fluid
theory. We assume that the plasma ions form an immobile neutralizing
background of density $n_0$ and define the plasma electron density $n$,
velocity ${\bf v}={\bf \beta }c$, and the plasma frequency $\omega _p=\left(
4\pi n_0e^2/m\right) ^{1/2}$. We now write the fluid equations for plasma
electrons in the presence of an ultrarelativistic beam with homogeneous
density $n_b$, length $d$, moving with the velocity ${\bf u}_b={\bf \beta }%
_bc$ along the $z$ axis and external constant magnetic field ${\bf B}_0$ ($%
B_{0x}=B_0\sin \alpha $, $B_{0y}=0$, $B_{0z}=B_0\cos \alpha $, where $\alpha 
$ is the angle between ${\bf B}_0$ and ${\bf u}_b$):

\begin{equation}
{\bf \nabla \times B}=\frac 1c\frac{\partial {\bf E}}{\partial t}-4\pi en%
{\bf \beta -}4\pi e{\bf \beta }_bn_b,
\end{equation}

\begin{equation}
{\bf \nabla \times E}=-\frac 1c\frac{\partial {\bf B}}{\partial t},\qquad 
{\bf \nabla }\cdot {\bf B}=0,
\end{equation}

\begin{equation}
{\bf \nabla }\cdot {\bf E}=-4\pi e\left( n-n_0\right) -4\pi en_b,
\end{equation}

\begin{equation}
\frac{\partial {\bf \rho }}{\partial t}+c\left( {\bf \beta \nabla }\right) 
{\bf \rho }=-\frac e{mc}\left[ {\bf E}+{\bf \beta }\times \left( {\bf B}+%
{\bf B}_0\right) \right] ,
\end{equation}

\begin{equation}
\frac{\partial n}{\partial t}+{\bf \nabla }\cdot \left( n{\bf v}\right) =0,
\end{equation}
where ${\bf \rho }={\bf \beta }/\left( 1-\beta ^2\right) ^{-1/2}$, $e$ is
the absolute value of electron charge.

If we assume that the wave motion is a function only of the variable $\xi
=z-ut$ (the wave ansatz, with $u_b$ taken as the wave phase velocity) we
obtain nonlinear differential equations for the dimensionless momentum and
field components:

\begin{equation}
n(\xi )=n_0\frac{\beta _b\sqrt{1+\rho ^2}}{\beta _b\sqrt{1+\rho ^2}-\rho _z},
\end{equation}

\begin{equation}
\frac{dE_{x;y}(\xi )}{d\xi }=\frac{mc\omega _p}e\frac{\beta _b^2\gamma _b^2}{%
\lambda _p}\frac{\rho _{x;y}}{\beta _b\sqrt{1+\rho ^2}-\rho _z},
\end{equation}

\begin{equation}
\frac{dE_z(\xi )}{d\xi }=-\frac{mc\omega _p}e\frac 1{\lambda _p}\left( \frac{%
\rho _z}{\beta _b\sqrt{1+\rho ^2}-\rho _z}+\frac{n_b}{n_0}\right) ,
\end{equation}

\begin{equation}
\frac{d\rho _x(\xi )}{d\xi }=\frac e{mc^2\beta _b}E_x(\xi )+\frac{\cos
\alpha }{\lambda _B}\frac{\rho _y}{\beta _b\sqrt{1+\rho ^2}-\rho _z},
\end{equation}

\begin{equation}
\frac{d\rho _y(\xi )}{d\xi }=\frac e{mc^2\beta _b}E_y(\xi )+\frac 1{\lambda
_B}\frac{\rho _z\sin \alpha -\rho _x\cos \alpha }{\beta _b\sqrt{1+\rho ^2}%
-\rho _z},
\end{equation}

\begin{equation}
\left( \beta _b\sqrt{1+\rho ^2}-\rho _z\right) \frac{d\rho _z}{d\xi }=\frac
e{mc^2}\left[ E_z\sqrt{1+\rho ^2}+\frac 1{\beta _b}\left( \rho _xE_x+\rho
_yE_y\right) \right] -\frac{\sin \alpha }{\lambda _B}\rho _y,
\end{equation}
where $\gamma _b^{-2}=1-\beta _b^2,$ $\lambda _p=c/\omega _p$, $\lambda
_B=c/\omega _B$, $\omega _B=eB_0/mc$. The components of induced magnetic
field are determined by means of relations: $E_x=\beta _bB_y$, $E_y=\beta
_bB_x$, $B_z=0$. Note that as it follows from Eqs. (6)-(11), $\rho _x=\rho
_y=E_x=E_y=0$ in the absence of external magnetic field (${\bf B}_0=0$) and
Eqs.(8) and (11) pass into the well known equations derived in Refs.[9] and
[15]. So, in an anisotropic magnetized plasma, besides the longitudinal
motion, the one-dimensional bunch excites the motion of plasma electrons in
the direction normal to that of bunch travel. When $\alpha =0$ (the magnetic
field is directed along ${\bf u}_b$) the magnetic field has no influence on
the excitation of wake waves and we arrive at the known equations of motion
that describe only the longitudinal motion of electrons.

To investigate the set of Eqs. (6)-(11) it is convenient to exclude the
electric field and obtain an equation containing the components of
dimensionless momentum:

\[
\frac{d^2}{d\xi ^2}\left[ \cos \alpha \sqrt{1+\rho ^2}-\beta _b\left( \rho
_x\sin \alpha +\rho _z\cos \alpha \right) \right] +
\]

\begin{equation}
+\frac 1{\lambda _p^2}\frac{\beta _b^2\gamma _b^2\rho _x\sin \alpha -\rho
_z\cos \alpha }{\beta _b\sqrt{1+\rho ^2}-\rho _z}=\frac{\cos \alpha }{%
\lambda _p^2}\frac{n_b}{n_0}.
\end{equation}
Then the component $\rho _y$ is eliminated from Eqs. (9) and (11). The
following equation is obtained:

\begin{eqnarray}
&&\frac d{d\xi }\left( \rho _x\sin \alpha +\rho _z\cos \alpha \right) \\
&=&\frac e{mc^2\beta _b}\cos \alpha \frac{\left( \beta _b\sqrt{1+\rho ^2}%
{\rm tg}\alpha +\rho _x-\rho _z{\rm tg}\alpha \right) E_x+\rho _yE_y+\beta _b%
\sqrt{1+\rho ^2}E_z}{\beta _b\sqrt{1+\rho ^2}-\rho _z}.  \nonumber
\end{eqnarray}

In what follows we shall assume the external magnetic field to be
comparatively strong so that the condition $\omega _B\gg \omega _p$ (or $%
\lambda _B\ll \lambda _p$) were observed. Hence we obtain the following
restriction on unperturbed concentration of plasma electrons:

\begin{equation}
B_0>3\times 10^{-6}\sqrt{n_0},
\end{equation}
where $n_0$ is measured in cm$^{-3}$, $B_0$ in kG. The condition (14) is
always observed in the range of parameters $n_0<10^{15}$cm$^{-3}$ and $%
B_0<100$ kG. In the zeroth-order approximation in $\lambda _B$ one obtains
from Eqs. (7) and (9) $\rho _y=E_y=B_x=0$, from Eq. (10) $\rho _z\sin \alpha
=\rho _x\cos \alpha $ $($or $\rho _z=\rho \cos \alpha ,$ $\rho _x=\rho \sin
\alpha )$, i.e., the plasma electrons move only in the direction of external
magnetic field. From Eqs. (12) and (13) we obtain in case of $\lambda
_B\rightarrow 0$:

\begin{eqnarray}
&&\frac{d^2}{d\xi ^2}\left( \cos \alpha \sqrt{1+\rho ^2}-\beta _b\rho
\right) + \\
+\frac{\gamma _b^2\sin ^2\alpha -1}{\lambda _p^2}\frac \rho {\beta _b\sqrt{%
1+\rho ^2}-\rho \cos \alpha } &=&\frac{\cos \alpha }{\lambda _p^2}\frac{n_b}{%
n_0},  \nonumber
\end{eqnarray}

\begin{equation}
\frac{\beta _b\sqrt{1+\rho ^2}-\rho \cos \alpha }{\sqrt{1+\rho ^2}}\frac{%
d\rho }{d\xi }=\frac e{mc^2}\left( E_x\sin \alpha +E_z\cos \alpha \right) .
\end{equation}

To express the components of electric field $E_x$ and $E_z$ through the
dimensionless momentum of electrons $\rho $, one is to obtain the second
equation that connects the components of the field. One can derive it from
Eqs. (4) and (5) by eliminating the momentum of electrons. As a result one
has

\begin{equation}
\frac d{d\xi }\left( E_x\cos \alpha +\beta _b^2\gamma _b^2E_z\sin \alpha
\right) =-\frac{mc\omega _p}e\beta _b^2\gamma _b^2\frac{\sin \alpha }{%
\lambda _p}\frac{n_b}{n_0}.
\end{equation}

Integrating the system of Eqs. (16) and (17) taking into account the
boundary conditions at the bunch fronts $\xi =d$ ($E_x(d)=E_z(d)=0$) and $%
\xi =0$ ($E_x(-0)=E_x(+0)$, $E_z(-0)=E_z(+0)$) one obtains the following
expressions for the field components $E_x$ and $E_z$ inside ($0\leq \xi \leq
d$, $n_b\neq 0$) and outside ($\xi <0$, $n_b=0$) the bunch:

$0\leq \xi \leq d;$

\begin{equation}
E_z(\xi )=-\frac{mc\omega _p}e\frac 1{1-\gamma _b^2\sin ^2\alpha }\left(
\Phi (\xi )\cos \alpha +\beta _b^2\gamma _b^2\sin ^2\alpha \frac{n_b}{n_0}%
\frac{d-\xi }{\lambda _p}\right) ,
\end{equation}

\begin{equation}
E_x(\xi )=\beta _bB_y(\xi )=\frac{mc\omega _p}e\frac{\beta _b^2\gamma
_b^2\sin \alpha }{1-\gamma _b^2\sin ^2\alpha }\left( \Phi (\xi )+\cos \alpha 
\frac{n_b}{n_0}\frac{d-\xi }{\lambda _p}\right) .
\end{equation}

$\xi <0;$

\begin{equation}
E_z(\xi )=-\frac{mc\omega _p}e\frac 1{1-\gamma _b^2\sin ^2\alpha }\left(
\Phi (\xi )\cos \alpha +\beta _b^2\gamma _b^2\sin ^2\alpha \frac{n_b}{n_0}%
\frac d{\lambda _p}\right) ,
\end{equation}

\begin{equation}
E_x(\xi )=\beta _bB_y(\xi )=\frac{mc\omega _p}e\frac{\beta _b^2\gamma
_b^2\sin \alpha }{1-\gamma _b^2\sin ^2\alpha }\left( \Phi (\xi )+\cos \alpha 
\frac{n_b}{n_0}\frac d{\lambda _p}\right) ,
\end{equation}
where

\begin{equation}
\Phi (\xi )=\lambda _p\frac d{d\xi }\left( \cos \alpha \sqrt{1+\rho ^2(\xi )}%
-\beta _b\rho (\xi )\right) .
\end{equation}

Note that from the continuity of components of the field and momentum of
electrons on the boundaries of the bunch there follows the continuity of
function $\Phi (\xi )$ at$\,\,\xi =0$ and $\xi =d$.

Thus, Eqs.(15) and (18)-(22) determine the wake wave in the magnetized
plasma for large value of the external magnetic field. For $\alpha =0$ we
obtain from Eqs. (18)-(22) that $E_x=B_y=0$, and Eqs. (15), (18) and (20)
coincide with equations given in Refs. [9, 15], that describe longitudinal
oscillations of the plasma electrons. When $\alpha $ has values different
from zero, the nature of solutions of Eqs. (15) and (18)-(22) is drastically
changed. First, as we shall see below, the behavior of solution of Eq. (15)
depends on the fact whether the angle $\alpha $ is greater or less than a
definite value $\Delta =\arcsin (1/\gamma _b)$. Second, owing to an
anisotropy of plasma due to the presence of an external magnetic field,
besides the longitudinal components there arise transverse components (as
well as an induced magnetic field). Furthermore, owing to the anisotropy of
plasma the components of the wake field (Eqs. (18)-(22)) contain the Coulomb
field of the bunch (the second terms in Eqs. (18)-(22)) that depends along
with other parameters also on the angle $\alpha $.

Below we shall analyze the Eq. (15) inside and behind the bunch; inside, on
and outside the anisotropy cone $\alpha =\Delta $.

\section{ELECTRON MOMENTUM INSIDE THE BUNCH FOR ARBITRARY VALUES OF THE
RATIO $n_B/n_0$}

Now let us consider the equation of plasma motion (15) inside the bunch. In
this range $0\leq \xi \leq d,\,\,n_b={\rm const}\neq 0$. The integration of
Eq. (15) with due regard for boundary conditions $\rho (d)=\Phi (d)=0$ will
give

\begin{equation}
\Phi (\xi )=\pm \sqrt{2}\left( A-A\sqrt{1+\rho ^2}-B\rho \right) ^{1/2},
\end{equation}
where

\begin{equation}
A=1-\gamma _b^2\sin ^2\alpha -\frac{n_b}{n_0}\cos ^2\alpha ,\quad B=\frac{n_b%
}{n_0}\beta _b\cos \alpha .
\end{equation}

In Eq. (23) one is to take the sign plus (minus) when $\rho (\xi )$
decreases (increases) with increasing $\xi $. As it follows from the Eq.
(23), for solution of Eq. (15) it is necessary that the radicand in (23)
assume only the positive values. Hence follows a restriction on the momentum
of plasma electrons: $\rho \leq 0$. Taking into account this restriction one
is to take the sign ''minus'' during the integration of Eq. (23). Making
allowance for boundary conditions at the front boundary of the bunch we
obtain from (22) and (23)

\begin{equation}
\frac{d-\xi }{\lambda _p}\sqrt{2}=\int_\rho ^0\frac{\left( \beta _b\sqrt{%
1+\eta ^2}-\eta \cos \alpha \right) d\eta }{\sqrt{1+\eta ^2}\sqrt{A-A\sqrt{%
1+\eta ^2}-B\eta }},
\end{equation}
that determines an implicit dependence of $\rho $ on $\xi $. Note that the
expression obtained in [9] will follow from (25) at $\alpha =0$.

To proceed with an analysis of the expression (25) one has to know whether $%
\alpha $ is greater or less than $\Delta $ and the dependence of $\alpha $
on the density of bunch.

First consider the case when $\alpha $ is inside the anisotropy cone ($%
\alpha <\Delta $).

\subsection{Inside the Cone: $\alpha <\Delta $}

Consider first the case of small densities of the bunch.

\subsubsection{The Case of $\frac{n_b}{n_0}<\gamma _b^2\frac{\cos \alpha
\,-\beta _b}{\cos \alpha }=b_1$}

In this case $0<B<A$ and one can easily show that the range of variation of $%
\rho $ is determined by the relation $-\rho _0\leq \rho \leq 0$, where

\begin{equation}
\rho _0=\frac{2\beta _b\varepsilon }{1-\beta _b^2\varepsilon ^2},\quad
\varepsilon =\frac{\frac{n_b}{n_0}\cos \alpha }{1-\gamma _b^2\sin ^2\alpha -%
\frac{n_b}{n_0}\cos ^2\alpha }.
\end{equation}
Then the dependence of $\rho $ on $\xi $ has a periodic behavior with
oscillation amplitude $\rho _0$.

The integration of (25) will give an implicit dependence of $\rho $ on $\xi $
for afore-mentioned values of the magnetic field orientation angle $\alpha $
and density of bunch:

\begin{eqnarray}
\frac{d-\xi }{\lambda _p} &=&\frac 1{1-\beta _b\varepsilon }\sqrt{\frac{%
2\left( 1+\varepsilon \cos \alpha \right) }{1-\gamma _b^2\sin ^2\alpha }}%
\left\{ \frac{\sqrt{2}\beta _b\left( 1+\varepsilon \cos \alpha \right) }{%
\sqrt{1+\beta _b\varepsilon }}E(\psi ,k)-\right. \\
&&\left. -\left( \beta _b+\cos \alpha \right) \sqrt{1-\beta _b\varepsilon
\rho -\sqrt{1+\rho ^2}}\right\} ,  \nonumber
\end{eqnarray}
where

\begin{equation}
\psi =\arcsin \left( \frac 1k\sqrt{1-\rho -\sqrt{1+\rho ^2}}\right) ,\quad k=%
\sqrt{\frac{2\beta _b\varepsilon }{1+\beta _b\varepsilon }}<1,
\end{equation}
$E(\psi ,k)$ is the elliptic integral of the second kind, $0\leq \beta
_b\varepsilon <1$. When $\rho =\rho _0$, assuming in Eq. (27) that $\xi =0$
one obtains the length of bunch $d_0$ or the wave half length inside the
bunch

\begin{equation}
\lambda _{{\rm in}}=2d_0=\frac{4\lambda _p\beta _b}{1-\beta _b\varepsilon }%
\sqrt{\frac{\left( 1+\varepsilon \cos \alpha \right) ^3}{\left( 1-\gamma
_b^2\sin ^2\alpha \right) \left( 1+\beta _b\varepsilon \right) }}E(k),
\end{equation}
where $E(k)$ is a complete elliptic integral of the second kind.

The dependence of wavelength inside the bunch on the bunch density is shown
in Fig.1 for values of $\alpha =0$ (solid line), $\alpha =0.2\Delta $
(dashed line), $\alpha =0.4\Delta $ (dotted line) and $0\leq n_b/n_0\leq b_1$%
.

For values of ratio $n_b/n_0\geq b_1$the only constraint on the region of
allowed values of $\rho $ is the condition $\rho \leq 0$. Therefore, the
dependence of $\rho $ on $\xi $ will get nonperiodic and $\rho (\xi )$ will
increase infinitely with $\xi $.

\subsubsection{The Case of $\frac{n_b}{n_0}=b_1$}

In this case $A=B$ and integration of expression (25) gives

\begin{eqnarray}
\frac{d-\xi }{\lambda _p} &=&\frac 12\sqrt{\frac{\left( \cos \alpha +\beta
_b\right) ^3}{2\beta _b\left( 1-\gamma _b^2\sin ^2\alpha \right) }}\times \\
&&\left\{ \sqrt{1-\rho -\sqrt{1+\rho ^2}}\left( \sqrt{1+\rho ^2}-\rho -2%
\frac{\cos \alpha -\beta _b}{\cos \alpha +\beta _b}\right) +\right. 
\nonumber \\
&&\left. +\ln \left[ \left( 1+\sqrt{1-\rho -\sqrt{1+\rho ^2}}\right) \sqrt{%
\sqrt{1+\rho ^2}-\rho }\right] \right\} .  \nonumber
\end{eqnarray}
One can see from Eq. (30) that at large values of $d-\xi \gg \lambda _p$,
i.e., with the distance from the front boundary of the bunch $\left| \rho
(\xi )\right| $ grows linearly depending on $d-\xi $.

\subsubsection{The Case of $b_1<\frac{n_b}{n_0}<\gamma _b^2\frac{\cos \alpha
+\beta _b}{\cos \alpha }=b_2$}

In this case $-B<A<B$. Taking into account this inequality on obtains from
Eq. (25)

\begin{eqnarray}
\frac{d-\xi }{\lambda _p} &=&\frac{\sqrt{2}}{B^2-A^2}\left\{ (B+|A|)\left(
\cos \alpha +\beta _b\right) \sqrt{A-A\sqrt{1+\rho ^2}-B\rho }+\right. \\
&&\left. +\left( B\cos \alpha +\beta _b|A|\right) \left[ \frac{B-|A|}{\sqrt{B%
}}F(\chi ,\kappa )-2\sqrt{B}E(\chi ,\kappa )\right] \right\} ,  \nonumber
\end{eqnarray}
where

\begin{equation}
\chi =\arccos \sqrt{\sqrt{1+\rho ^2}+\rho },\quad \kappa =\sqrt{\frac{B+|A|}{%
2B}},
\end{equation}
$F(\chi ,\kappa )$ is an elliptic integral of the first kind , $\kappa <1.$

\subsubsection{The Case of $\frac{n_b}{n_0}=b_2$}

In this case $A=-B$. One obtains from the general form of Eq. (25)

\begin{eqnarray}
\frac{d-\xi }{\lambda _p} &=&\frac 1{\sqrt{2B}}\left\{ \left[ \cos \alpha
+\beta _b-\frac 12\left( \cos \alpha -\beta _b\right) \left( \sqrt{1+\rho ^2}%
+\rho \right) \right] \times \right.  \\
&&\left. \times \sqrt{\sqrt{1+\rho ^2}-\rho -1}-\frac 12\left( \cos \alpha
-\beta _b\right) {\rm arctg}\sqrt{\sqrt{1+\rho ^2}-\rho -1}\right\} . 
\nonumber
\end{eqnarray}

\subsubsection{The Case of $\frac{n_b}{n_0}>b_2$}

Here $A<-B$. The calculation of the integral in Eq. (25) gives

\begin{equation}
\frac{d-\xi }{\lambda _p}=\frac{\sqrt{2}}{A^2-B^2}\left\{ \sqrt{A-A\sqrt{%
1+\rho ^2}-B\rho }\right. \times 
\end{equation}

\begin{eqnarray}
&&\times \left[ |A|\cos \alpha -\beta _bB+\left( \beta _b|A|-B\cos \alpha
\right) \frac{\sqrt{A^2-B^2}-A\sqrt{1+\rho ^2}-B\rho }{B\sqrt{1+\rho ^2}%
+A\rho }\right] +  \nonumber \\
&&\left. +\left( B\cos \alpha +\beta _bA\right) \left[ \sqrt{\sqrt{A^2-B^2}-A%
}E(\sigma ,r)+\frac A{\sqrt{\sqrt{A^2-B^2}-A}}F(\sigma ,r)\right] \right\}  
\nonumber
\end{eqnarray}
where

\begin{equation}
\sigma =\arcsin \sqrt{\frac{A\sqrt{1+\rho ^2}+B\rho -A}{A\sqrt{1+\rho ^2}%
+B\rho +\sqrt{A^2-B^2}}},\quad r=\sqrt{\frac{2\sqrt{A^2-B^2}}{\sqrt{A^2-B^2}%
-A}}.
\end{equation}

In all cases under consideration in items 3-5, at large values of $d-\xi \gg
\lambda _p$ \thinspace \thinspace $\left| \rho (\xi )\right| $ is a
quadratic function of the distance to the front boundary of the bunch $d-\xi 
$. At $\alpha =0$ the expressions obtained in items 1-5 coincide with the
results given in Ref. [9].

\subsection{On \thinspace the Cone: $\alpha =\Delta $}

In case of $\alpha =\Delta $ the Eq. (15) for the momentum of electrons is
simplified ant takes the following form:

\begin{equation}
\frac{d^2}{d\xi ^2}\left( \sqrt{1+\rho ^2}-\rho \right) =\frac 1{\lambda
_p^2}\frac{n_b}{n_0}.
\end{equation}
The integration of Eq. (36) with due regard for boundary conditions gives
the following expression for $\rho (\xi )$:

\begin{equation}
\rho (\xi )=-\frac{\frac{n_b}{2n_0}\left( \frac{d-\xi }{\lambda _p}\right)
^2\left[ \frac{n_b}{2n_0}\left( \frac{d-\xi }{\lambda _p}\right) ^2+2\right] 
}{2\left[ \frac{n_b}{2n_0}\left( \frac{d-\xi }{\lambda _p}\right)
^2+1\right] }.
\end{equation}

As it follows from Eq. (37) for arbitrary densities of the bunch the
momentum of electrons increases monotonically with the distance from the
front boundary of the bunch.

\subsection{Outside \thinspace the Cone: $\alpha >\Delta $}

At large values of angle $\alpha $ ($\alpha >\Delta $) the value of A is
always negative. As follows from an analysis of Eq. (23), in such a case $%
\rho $ is a monotonic function of $\xi $ and is described by Eqs. (31)-(35)
for different values of the bunch density. So, when the conditions $%
n_b/n_0<b_2$, $n_b/n_0=b_2$ and $n_b/n_0>b_2$ are approached, the implicit
dependences of $\rho $ on $\xi $ are given by Eqs. (34), (33) and (31)
respectively.

\section{THE FIELDS INSIDE THE BUNCH FOR ARBITRARY VALUES OF THE RATIO $%
n_b/n_0$}

In this Section the Eqs. (18) and (19) describing the induced
electromagnetic fields inside the bunch were analyzed. Consider at first the
case $\alpha \neq \Delta $. We shall begin with an analysis of a transverse
component of electric field $E_x$. In the previous Section it was shown that
the momentum of electrons inside the bunch meets the condition $\rho (\xi
)\leq 0$. As follows from this inequality, Eq. (7) and the boundary
condition $E_x(d)=0$, inside the bunch $E_x$ is positive and monotonically
increases with distance from the front boundary of the bunch. The function $%
E_x$ (as well as $B_y$) takes on the maximum value at the rear boundary of
the bunch, at $\xi =0$. From Eq. (18) one obtains for the maximum value of $%
E_x$:

\begin{equation}
E_{x\max }=\frac{mc\omega _p}e\frac{\beta _b^2\gamma _b^2\sin \alpha }{%
1-\gamma _b^2\sin ^2\alpha }\left( \pm 2^{1/2}\sqrt{A-A\sqrt{1+\rho ^2(0)}%
-B\rho (0)}+\frac{n_b}{n_0}\frac d{\lambda _p}\cos \alpha \right) .
\end{equation}

In Eq. (38) one takes the sign ''plus'' (''minus'') if $\rho (\xi )$ is
decreased (increased) with increasing of $\xi $. In particular, if $%
n_b/n_0<b_1$ (the momentum of electrons inside the bunch is periodically
changing with $\xi $ and is given by Eq. (27)) and $d=\lambda _{in}/2$, then 
$\rho (0)=-\rho _0$ and the maximum value of $E_x$ as obtained from (38) is

\begin{equation}
E_{x\max }=\frac{mc\omega _p}e\frac{n_b}{n_0}\frac{\lambda _{{\rm in}}}{%
2\lambda _p}\frac{\beta _b^2\gamma _b^2\sin \alpha \cos \alpha }{1-\gamma
_b^2\sin ^2\alpha }.
\end{equation}

In all other cases discussed in the previous Section, $E_{x\max }$ is
determined by the Eq. (38) with the sign ''minus''.

The longitudinal component $E_z$ of the field satisfies the Eq. (8). In case
of $n_b/n_0\geq \cos \alpha /(\beta _b+\cos \alpha )\equiv b_c$, $dE_z/d\xi
<0$ and $E_z(\xi )$ is positive and monotonically increase with the distance
from the front boundary of the bunch. In case of $n_b/n_0<b_c$ the
longitudinal field $E_z$ first monotonically increases with the distance
from the front boundary of the bunch, than reaching the maximum at $\rho (%
\widetilde{\xi })=-\widetilde{\rho },$ where

\begin{equation}
\widetilde{\rho }=\frac{\beta _b\theta }{\sqrt{\cos ^2\alpha -\beta
_b^2\theta ^2}},\quad \theta =\frac{n_b/n_0}{1-n_b/n_0},
\end{equation}
monotonically decreases to the rear boundary ($\xi =0$).

The maximum value of longitudinal field $E_z$ in the point $\widetilde{\xi }$
follows from expression (18), where the substitutions $\rho =-\widetilde{%
\rho }$ and $\xi =\widetilde{\xi }$ are made:

\begin{equation}
\widetilde{E}_z=\frac{mc\omega _p/e}{1-\gamma _b^2\sin ^2\alpha }\left(
2^{1/2}\cos \alpha \sqrt{A-A\sqrt{1+\widetilde{\rho }^2}+B\widetilde{\rho }}%
-\beta _b^2\gamma _b^2\sin ^2\alpha \frac{n_b}{n_0}\frac{d-\widetilde{\xi }}{%
\lambda _p}\right) ,
\end{equation}
where the value of $(d-\widetilde{\xi })/\lambda _p$ is expressed through $%
\widetilde{\rho }$ by means of expressions (27), (30), (31), (33) and (34).
The value of longitudinal field at the rear boundary of the bunch ($\xi =0$)
follows from the Eq. (18):

\begin{equation}
E_z(0)=\frac{-mc\omega _p/e}{1-\gamma _b^2\sin ^2\alpha }\left( \pm
2^{1/2}\cos \alpha \sqrt{A-A\sqrt{1+\rho ^2(0)}-B\rho (0)}+\beta _b^2\gamma
_b^2\sin ^2\alpha \frac{n_b}{n_0}\frac d{\lambda _p}\right) .
\end{equation}

In the general case the maximum value of longitudinal field inside the bunch
is

\begin{equation}
E_{z\max }=\max \left[ \left| E_z(0)\right| ;\quad \widetilde{E}_z\right] .
\end{equation}

In particular, if $n_b/n_0<b_1$ and $d=\lambda _{{\rm in}}/2$ the value of $%
\widetilde{E}_z$ is determined by the Eq. (41), and for $E_z(0)$ we obtain
from (18), (23), (26) and (29):

\begin{equation}
E_z(0)=-\frac{mc\omega _p}e\frac{\beta _b^2\gamma _b^2\sin ^2\alpha }{%
1-\gamma _b^2\sin ^2\alpha }\frac{n_b}{n_0}\frac{\lambda _{{\rm in}}}{%
2\lambda _p}<0.
\end{equation}

The case $\cos \alpha =\beta _b$ $\,$(or $\gamma _b\sin \alpha =1$) is to be
considered separately, as in this case the Eq. (18), (19) make no sense.
Indeed, it follows from expressions (22) and (36) that both the denominator
and the determinant of the expressions (18) and (19) tend to zero when $%
\alpha \rightarrow \Delta $. The electromagnetic field inside the bunch in
this case may be found from Eqs. (7), (8) (or Eqs. (7) and (17)) and (37).

As a result we obtain

\begin{equation}
E_x(\xi )=\frac{mc\omega _p}e\frac{\beta _b\gamma _b}4\left[ \frac{d-\xi }{%
\lambda _p}\frac{\frac{n_b}{n_0}\left( \frac{d-\xi }{\lambda _p}\right) ^2+1%
}{\frac{n_b}{2n_0}\left( \frac{d-\xi }{\lambda _p}\right) ^2+1}-\sqrt{\frac
2{(n_b/n_0)}}\arctan \left( \sqrt{\frac{n_b}{2n_0}}\frac{d-\xi }{\lambda _p}%
\right) \right] ,
\end{equation}

\begin{eqnarray}
E_z(\xi ) &=&\frac 14\frac{mc\omega _p}e\left[ \frac{d-\xi }{\lambda _p}%
\frac{\frac{n_b}{n_0}(\frac{2n_b}{n_0}-1)\left( \frac{d-\xi }{\lambda _p}%
\right) ^2+\frac{4n_b}{n_0}-1}{\frac{n_b}{2n_0}\left( \frac{d-\xi }{\lambda
_p}\right) ^2+1}+\right.  \\
&&\left. +\sqrt{\frac 2{(n_b/n_0)}}\arctan \left( \sqrt{\frac{n_b}{2n_0}}%
\frac{d-\xi }{\lambda _p}\right) \right] .  \nonumber
\end{eqnarray}

The values $\widetilde{\rho }$, $\widetilde{\xi }$ and $\widetilde{E}_z$ are
determine by means of the following expressions:

\begin{equation}
\widetilde{\rho }=\frac \theta {\sqrt{1-\theta ^2}},
\end{equation}

\begin{equation}
\frac{d-\widetilde{\xi }}{\lambda _p}=\sqrt{\frac 2{(n_b/n_0)}\left( \frac 1{%
\sqrt{1-\frac{2n_b}{n_0}}}-1\right) },
\end{equation}

\begin{eqnarray}
\widetilde{E}_z &=&\frac 14\frac{mc\omega _p}e\sqrt{\frac 2{(n_b/n_0)}}%
\left[ \sqrt{\frac 1{\sqrt{1-\frac{2n_b}{n_0}}}-1}\left( \sqrt{1-\frac{2n_b}{%
n_0}}-2(1-\frac{2n_b}{n_0})\right) +\right.  \\
&&\left. +\arctan \sqrt{\frac 1{\sqrt{1-\frac{2n_b}{n_0}}}-1}\right] . 
\nonumber
\end{eqnarray}

Below we shall consider the equation of motion of plasma (15) outside the
bunch ($\xi <0$).

\section{THE\ WAKE\ FIELD}

\subsection{Inside the Cone: $\alpha <\Delta $}

To find the wake field $E_x(\xi )$ and $E_z(\xi )$ behind the bunch ($\xi
\leq 0$), one has to integrate Eq. (15) for $n_b=0$ taking into account the
condition of continuity of fields and momentum $\rho $ at the rear boundary
of the bunch $\xi =0$. In this case the fields are determined by means of
expressions (20)-(22), where now

\begin{equation}
\Phi (\rho )=\pm \sqrt{2\left( 1-\gamma _b^2\sin ^2\alpha \right) \left( 
\sqrt{1+\rho _{\max }^2}-\sqrt{1+\rho ^2}\right) }.
\end{equation}
Here

\begin{equation}
\rho _{\max }=\sqrt{\left[ \frac{A-B\rho (0)}{1-\gamma _b^2\sin ^2\alpha }+%
\sqrt{1+\rho ^2(0)}\left( 1-\frac A{1-\gamma _b^2\sin ^2\alpha }\right)
\right] ^2-1}
\end{equation}
is the maximum allowable value of the momentum $\rho (\xi )$. As is seen
from expression (50), the momentum $\rho (\xi )$ (hence, the fields $E_x(\xi
)$, $E_z(\xi )$ and function $\Phi (\xi )$) is a periodic function and
changes in the range $-\rho _{\max }\leq \rho (\xi )\leq \rho _{\max }$. The
wake wavelength is determined from expressions (22), (50) and has the
following form

\begin{equation}
\lambda _{{\rm out}}=\frac{4\sqrt{2}\beta _b\lambda _p}{\sqrt{1-\gamma
_b^2\sin ^2\alpha }}\sqrt{\frac 2{1-p^2}}\left[ E(p)-\frac{1-p^2}%
2K(p)\right] ,
\end{equation}
where

\begin{equation}
p=\sqrt{\frac{\sqrt{1+\rho _{\max }^2}-1}{\sqrt{1+\rho _{\max }^2}+1}}.
\end{equation}

The dependence of wake wavelength ($\lambda _{{\rm out}}$) on its amplitude $%
\rho _{\max }$ is shown in Fig.2. It is seen that the wavelength increases
with the amplitude and angle $\alpha $.

As it follows from Eqs. (20)-(22) and (50), (51) the transverse ($E_x(\xi )$%
) and longitudinal ($E_z(\xi )$) wake waves change within the ranges ${\cal E%
}_{1x}\leq E_x(\xi )\leq {\cal E}_{2x}$, ${\cal E}_{1z}\leq E_z(\xi )\leq 
{\cal E}_{2z}$, respectively, where

\begin{eqnarray}
\left( 
\begin{tabular}{l}
${\cal E}_{1x}$ \\ 
${\cal E}_{2x}$%
\end{tabular}
\right) &=&\frac{mc\omega _p}e\frac{\beta _b^2\gamma _b^2\sin \alpha }{%
1-\gamma _b^2\sin ^2\alpha }\left[ \cos \alpha \frac{n_b}{n_0}\frac
d{\lambda _p}\mp \right. \\
&&\left. \left( \frac{2n_b}{n_0}\cos \alpha \right) ^{1/2}\sqrt{\cos \alpha
\left( \sqrt{1+\rho ^2(0)}-1\right) -\beta _b\rho (0)}\right] ,  \nonumber
\end{eqnarray}

\begin{eqnarray}
\left( 
\begin{tabular}{l}
${\cal E}_{1z}$ \\ 
${\cal E}_{2z}$%
\end{tabular}
\right) &=&-\frac{mc\omega _p/e}{1-\gamma _b^2\sin ^2\alpha }\left[ \beta
_b^2\gamma _b^2\sin ^2\alpha \frac{n_b}{n_0}\frac d{\lambda _p}\pm \right. \\
&&\left. \left( \frac{2n_b}{n_0}\cos ^3\alpha \right) ^{1/2}\sqrt{\cos
\alpha \left( \sqrt{1+\rho ^2(0)}-1\right) -\beta _b\rho (0)}\right] . 
\nonumber
\end{eqnarray}

It is seen from obtained expressions that $\left| {\cal E}_{1x}\right| <%
{\cal E}_{2x}$ and $\left| {\cal E}_{1z}\right| >\left| {\cal E}_{2z}\right| 
$. Thus, the maximum values of transverse and longitudinal fields behind the
bunch are ${\cal E}_{2x}$ and $\left| {\cal E}_{1z}\right| $ respectively.
Here the value of transverse field varies with respect to some positive
quantity (the first term in Eq. (54)), and the value of longitudinal field
varies with respect to some negative quantity (the first term in Eq. (55)).

Now consider the efficiency of wake field excitation for different values of
the density and thickness of the bunch. If the momentum of plasma electrons
inside the bunch is a periodic function (the bunch density meets the
condition of subsection A1 of Section III, $n_b/n_0<b_1$), then inside the
bunch the variation range of $\rho $ is defined by the relation $-\rho
_0\leq \rho \leq 0$, where $\rho _0$ is determined by expression (26). In
case when an integer number of wavelengths is interposed within the bunch ($%
d=(n+1)\lambda _{{\rm in}}$, where $n=0,~1,~2~...$), then it follows from
results of subsection A1 of the Section III that $\rho (0)=0$, and in virtue
of expression (51) $\rho _{\max }=0$. In this case (as in the absence of
magnetic field [9]) the momentum of electrons for $\xi \leq 0$ is zero, and
the density of plasma is coincident with the equilibrium value $n_0$. As
follows from expressions (20), (21), and (50) and (51), the components of
electric fields for $\xi \leq 0$ are constant and are determined by the
second terms of expressions (20) and (21). Thus, when the condition of $%
d=(n+1)\lambda _{{\rm in}}$ is observed in an anisotropic plasma, a constant
electric field may arise behind the bunch that is perpendicular to the
external magnetic field according to expressions (54) and (55).

In case of $d=(n+1/2)\lambda _{{\rm in}}$, the amplitude of wake field
oscillations is maximum $\rho _{\max }=\left| \rho (0)\right| =\rho _0$. The
maximum values of electric field components are

\begin{equation}
E_{z\max }^{({\rm out})}=\frac{mc\omega _p}e\beta _b\varepsilon \left[ \frac{%
2\cos \alpha }{\sqrt{\left( 1-\gamma _b^2\sin ^2\alpha \right) \left(
1-\beta _b^2\varepsilon ^2\right) }}+\frac{(n+1/2)\lambda _{{\rm in}}}{%
\lambda _p}\frac{\beta _b\gamma _b^2\sin ^2\alpha }{\cos \alpha \left(
1+\varepsilon \cos \alpha \right) }\right] ,
\end{equation}

\begin{equation}
E_{x\max }^{({\rm out})}=\frac{mc\omega _p}e\varepsilon \beta _b^2\gamma
_b^2\sin \alpha \left[ \frac{2\beta _b}{\sqrt{\left( 1-\gamma _b^2\sin
^2\alpha \right) \left( 1-\beta _b^2\varepsilon ^2\right) }}+\frac{\lambda _{%
{\rm in}}}{\lambda _p}\frac{n+1/2}{1+\varepsilon \cos \alpha }.\right]
\end{equation}

The condition of positivity of plasma density $n(\xi )$ (Eq. (6)) imposes a
restriction on possible values of the momentum of plasma electrons at $\xi
<0 $

\begin{equation}
-\infty <\rho (\xi )\leq \rho _{\max }\leq \frac{\beta _b\gamma _b}{\sqrt{%
1-\gamma _b^2\sin ^2\alpha }},
\end{equation}
that, in its turn, limits the values of electron momentum at the rear
boundary of the bunch $(\xi =0)$. For example, for the value of bunch
density $n_b/n_0<b_1$ Eq. (58) yields

\begin{equation}
-\rho _{\max }(0)\equiv -\frac{\beta _b\gamma _b}{\sqrt{1-\gamma _b^2\sin
^2\alpha }}\leq \rho (0)\leq 0.
\end{equation}

For a bunch, the density of which satisfies the condition $n_b/n_0<b_1$, the
electron momentum inside the bunch is a periodical function. In this case
from expressions (26) and (59) we obtain a restriction on the density of
bunch, that is more rigid than the condition $n_b/n_0<b_1$:

\begin{equation}
\frac{n_b}{n_0}\leq \frac{1-\gamma _b^2\sin ^2\alpha }{\cos \alpha \left(
2\cos \alpha +\sqrt{\gamma _b^{-2}-\sin ^2\alpha }\right) }=b_0<b_1.
\end{equation}

For the density of bunch $n_b/n_0=b_0$, the maximum values of fields have
the following forms:

\begin{eqnarray}
E_{z\max }^{({\rm out})} &=&\frac{mc\omega _p}e\sqrt{\gamma _b^2-1}\left[ 
\frac{\sqrt{2}\cos \alpha }{\sqrt{\left( 1-\gamma _b^2\sin ^2\alpha \right)
^{3/2}\left( \gamma _b\cos \alpha +\sqrt{1-\gamma _b^2\sin ^2\alpha }\right) 
}}+\right.  \\
&&\left. +\frac{\beta _b\gamma _b^2\sin ^2\alpha \left( \lambda _{{\rm in}%
}/\lambda _p\right) (n+1/2)}{\cos \alpha \left( 2\cos \alpha +\sqrt{\gamma
_b^{-2}-\sin ^2\alpha }\right) }\right] ,  \nonumber
\end{eqnarray}

\begin{eqnarray}
E_{x\max }^{({\rm out})} &=&\frac{mc\omega _p}e\beta _b^2\gamma _b^3\sin
\alpha \left[ \frac{\sqrt{2}\beta _b}{\sqrt{\left( 1-\gamma _b^2\sin
^2\alpha \right) ^{3/2}\left( \gamma _b\cos \alpha +\sqrt{1-\gamma _b^2\sin
^2\alpha }\right) }}+\right. \\
&&\left. +\frac{\left( \lambda _{{\rm in}}/\lambda _p\right) (n+1/2)}{%
2\gamma _b\cos \alpha +\sqrt{1-\gamma _b^2\sin ^2\alpha }}\right] . 
\nonumber
\end{eqnarray}

The dependence of wake waves on $\xi $ is shown in Figs. 3-6 for the values $%
\gamma _b=60$, $d=\lambda _{{\rm in}}/2$, $n_b/n_0=b_0(\alpha )$ (Figs. 3
and 4) and $n_b/n_0=0.8b_0(\alpha )$ (Figs. 5 and 6). Note that in Fig. 3
the curve calculated for the value $\alpha =0$ coincides with the result
obtained in Ref. [9].

In case of even higher densities of the bunch $n_b/n_0>b_0$, there is also a
limit on the thickness of the bunch $d\leq d_{\max }$, where $d_{\max }$ is
determined from expression (25) if in the latter we put $\xi =0$ and $\rho
=-\rho _{\max }(0)$:

\begin{equation}
d_{\max }=\frac{\lambda _p}{\sqrt{2}}\int_0^{\rho _{\max }(0)}\frac{\left(
\beta _b\sqrt{1+\eta ^2}+\eta \cos \alpha \right) d\eta }{\sqrt{1+\eta ^2}%
\sqrt{A-A\sqrt{1+\eta ^2}+B\eta }}.
\end{equation}
When $b_0<n_b/n_0<b_1$, the thickness of the bunch should satisfy the
condition $0\leq d\leq d_{\max }$ or $n\lambda _{{\rm in}}-d_{\max }\leq
d\leq n\lambda _{{\rm in}}+d_{\max }$, where $n=1,~2,~3,~...$ Note that in
this case $d_{\max }<\lambda _{{\rm in}}/2$. For $n_b/n_0\geq b_1$ the
thickness of the bunch satisfies only the condition $0\leq d\leq d_{\max }$.

In Fig. 7 the dependence of maximum thickness of the bunch its density is
shown at $n_b/n_0>b_0$ for $\alpha =0$ (solid line), $\alpha =0.4\Delta $
(dashed line), $\alpha =0.8\Delta $ (dotted line). As was expected, $d_{\max
}$ decreases with increasing density of the bunch.

Consider now the variation range of the value of $\,\rho (0)$ for$%
\,\,\,n_b/n_0>b_0$. From expressions (51) and (52) one has $-\rho _{\max
}(0)\leq \rho (0)\leq 0$, where $\rho _{\max }(0)$ is now a function of the
density of beam:

\begin{equation}
\rho _{\max }(0)=\frac{\gamma _b^2}{1-\gamma _b^2\sin ^2\alpha }\left[ \cos
\alpha \sqrt{G^2-\left( \gamma _b^{-2}-\sin ^2\alpha \right) }-\beta
_bG\right] ,
\end{equation}
where

\begin{equation}
G=\cos \alpha +\frac{\sqrt{1-\gamma _b^2\sin ^2\alpha }\left( \gamma _b\cos
\alpha -\sqrt{1-\gamma _b^2\sin ^2\alpha }\right) }{\left( n_b/n_0\right)
\cos \alpha }.
\end{equation}

One can find the values of electric fields at the rear boundary of the bunch
($\xi =0$) for $d=d_{\max }$ from expressions (18), (19), and (23), if one
makes a substitution $\rho =-\rho _{\max }(0)$ in the latter, where $\rho
_{\max }(0)$ is determined by the expression (64). As a result we have for
the value of $\Phi (0)$

\begin{equation}
\Phi (0)=-\sqrt{2}\left\{ \gamma _b\cos \alpha \sqrt{1-\gamma _b^2\sin
^2\alpha }-\gamma _b^2\left( G\cos \alpha -\beta _b\sqrt{G^2-\left( \gamma
_b^{-2}-\sin ^2\alpha \right) }\right) \right\} ^{1/2}.
\end{equation}

The maximum values of the longitudinal and transverse electric fields inside
the bunch are determined by expressions (41), (43) and (66).

\subsection{On the Cone: $\alpha =\Delta $}

In case of $\alpha =\Delta $ and $n_b=0$ the Eq. (36) takes the following
form:

\begin{equation}
\frac{d^2}{d\xi ^2}\left( \sqrt{1+\rho ^2}-\rho \right) =0.
\end{equation}

The integration of Eq. (67) with regard to the boundary conditions gives the
following expression for the momentum of electrons:

\begin{equation}
\rho (\xi )=-\frac{n_b}{2n_0}\frac d{\lambda _p}\frac{\frac{d/2-\xi }{%
\lambda _p}\left[ \frac{n_b}{n_0}\frac d{\lambda _p}\frac{d/2-\xi }{\lambda
_p}+2\right] }{\frac{n_b}{n_0}\frac d{\lambda _p}\frac{d/2-\xi }{\lambda _p}%
+1}.
\end{equation}

The components of electric field are determined from expressions (7), (8)
and (68)

\begin{equation}
E_x(\xi )=E_x(0)-\frac{mc\omega _p}e\frac{\beta _b\gamma _b}{2\lambda _p}\xi 
\frac{\frac{n_b}{n_0}\frac d{\lambda _p}}{\frac{n_b}{n_0}\frac d{\lambda _p}%
\frac{d/2-\xi }{\lambda _p}+1}\left\{ \frac{\frac d{\lambda _p}\left[ \frac{%
n_b}{2n_0}\left( \frac d{\lambda _p}\right) ^2+2\right] }{\frac{n_b}{n_0}%
\left( \frac d{\lambda _p}\right) ^2+2}-\frac \xi {\lambda _p}\right\} ,
\end{equation}

\begin{equation}
E_z(\xi )=E_z(0)+\frac{mc\omega _p}e\frac \xi {2\lambda _p}\frac{\frac{n_b}{%
n_0}\frac d{\lambda _p}}{\frac{n_b}{n_0}\frac d{\lambda _p}\frac{d/2-\xi }{%
\lambda _p}+1}\left\{ \frac{\frac d{\lambda _p}\left[ \frac{n_b}{2n_0}\left(
\frac d{\lambda _p}\right) ^2+2\right] }{\frac{n_b}{n_0}\left( \frac
d{\lambda _p}\right) ^2+2}-\frac \xi {\lambda _p}\right\} ,
\end{equation}
where $E_x(0)$ and $E_z(0)$ are the values of electric fields at the rear
boundary of the bunch, and can be determined from expressions (45) and (46).
As follows from obtained Eqs. (68)-(70), the modulo of momentum and electric
field grow linearly with the distance from the rear boundary of the bunch.

\subsection{Outside the Cone: $\alpha >\Delta $}

If the orientation angle of the magnetic field is outside the anisotropy
cone, then the dependence of the momentum of electrons and of electric
fields on coordinates is monotone.

Introducing the notations

\begin{eqnarray}
D &=&\gamma _b^2\sin ^2\alpha -1>0, \\
C &=&\frac{n_b}{n_0}\cos \alpha \left[ \cos \alpha \left( \sqrt{1+\rho ^2(0)}%
-1\right) -\beta _b\rho (0)\right] -D  \nonumber
\end{eqnarray}
and integrating the equations of motion we obtain an implicit dependence of
the momentum of electrons on $\xi $:

When $C>D>0;$

\begin{eqnarray}
-\frac \xi {\lambda _p} &=&\frac{\sqrt{2}\cos \alpha }D\left( \sqrt{C+D\sqrt{%
1+\rho ^2}}-\sqrt{C+D\sqrt{1+\rho ^2(0)}}\right) + \\
&&+\beta _b\sqrt{\frac 2D}\left\{ \sqrt{1+\frac CD}\left[ F\left( \nu \left(
\rho \right) ,\tau \right) -F\left( \nu _0,\tau \right) +E\left( \nu _0,\tau
\right) -E\left( \nu \left( \rho \right) ,\tau \right) \right] -\right. 
\nonumber \\
&&-\frac C{\sqrt{D(C+D)}}\left[ F\left( \mu \left( \rho \right) ,\tau
\right) -F\left( \mu _0,\tau \right) \right] +  \nonumber \\
&&\left. \frac{\rho (0)\sqrt{\sqrt{1+\rho ^2(0)}+\frac CD}}{\sqrt{1+\rho
^2(0)}+1}-\frac{\rho \sqrt{\sqrt{1+\rho ^2}+\frac CD}}{\sqrt{1+\rho ^2}+1}%
\right\} ,  \nonumber
\end{eqnarray}
where

\begin{eqnarray}
\tau &=&\sqrt{\frac{C-D}{C+D}},\quad \nu (\rho )=\arcsin \sqrt{\frac{C+D}{C+D%
\sqrt{1+\rho ^2}}},\quad \nu _0=\nu (\rho (0)), \\
\mu (\rho ) &=&\arcsin \frac{-\rho }{\sqrt{1+\rho ^2}+1},\quad \mu _0=\mu
(\rho (0)).  \nonumber
\end{eqnarray}

When $-D\leq C\leq D;$

\begin{eqnarray}
-\frac \xi {\lambda _p} &=&\frac{\sqrt{2}\cos \alpha }D\left( \sqrt{C+D\sqrt{%
1+\rho ^2}}-\sqrt{C+D\sqrt{1+\rho ^2(0)}}\right) + \\
&&+\frac{\beta _b}{\sqrt{D}}\left\{ F\left( \frac \pi 2-\nu \left( \rho
\right) ,\vartheta \right) -F\left( \frac \pi 2-\nu _0,\vartheta \right)
-\right.   \nonumber \\
&&-2\left[ E\left( \frac \pi 2-\nu (\rho ),\vartheta \right) -E\left( \frac
\pi 2-\nu _0,\vartheta \right) \right] +  \nonumber \\
&&\left. +\frac{\rho (0)\sqrt{2D}}{\sqrt{C+D\sqrt{1+\rho ^2(0)}}}-\frac{\rho 
\sqrt{2D}}{\sqrt{C+D\sqrt{1+\rho ^2}}}\right\} ,  \nonumber
\end{eqnarray}

where

\begin{equation}
\tau =\sqrt{\frac 12\left( 1-\frac CD\right) }.
\end{equation}

One can obtain the electric fields from expressions (20) and (21), where $%
\Phi $ is determined from the expression:

\begin{equation}
\Phi (\xi )=-\sqrt{2\left( \gamma _b^2\sin ^2\alpha -1\right) \left( \sqrt{%
1+\rho ^2(\xi )}+\frac CD\right) }.
\end{equation}

Taking into account the fact that in the space $-\infty <\xi \leq 0$ the
function $\rho (\xi )$ is monotonically increasing, the sign chosen in the
expression (76) is ''minus''.

\section{THE\ TRANSFORMER\ RATIO}

In this Section we shall consider the transformer ratio $R$ that is defined
as the ratio of maximum value of longitudinal wake field to that of
decelerating field inside the bunch $R=E_{z\,\max }^{({\rm out})}/E_{z\,\max
}^{({\rm in})}$ [4-9]. Since the periodic wake field is excited only in case
of $\alpha <\Delta $, below we shall consider the transformer ratio
precisely for these values of angle $\alpha $.

We would briefly remind the results obtained for $R$ in the absence of
magnetic field. As follows from the results of Refs. [4-9], the transformer
ratio monotonically increases with density of bunch, the maximum value $%
R_{\max }\simeq \sqrt{2\left( \gamma _b-1\right) }$ being achieved at $%
n_b/n_0\simeq 1/(2+1/\gamma _b)$. In the linear case ($n_b/n_0\ll 1)$ the
function $R$ assumes the value $R=2$. When $n_b/n_0>1/(2+1/\gamma _b)$, the
transformer ratio is monotonically decreasing and tends to unity, $R\simeq 1$%
, at very large values of bunch density. So, in this case for excitation of
wake waves the optimal is a bunch with length $d\leq d_{\max }\equiv \lambda
_{{\rm in}}/2$ and density $1/(2+1/\gamma _b)\leq n_b/n_0\leq 1/(1+\beta _b)$%
, for which the amplitude of wake waves and the transformer ratio take
maximum values.

Consider now the influence of external strong magnetic field on the
transformer ratio. In case of bunches of low density, when $n_b/n_0\leq b_0$%
, the maximum value of longitudinal wake field is determined from Eq. (55),
where $d=(n+1/2)\lambda _{{\rm in}}$, $(n=0,~1,~2,~...)$ and $\rho (0)=-\rho
_0$. One can determine the maximum decelerating field inside the bunch from
expressions (40) and (41). For very small values of the density of bunch ($%
n_b/n_0\ll b_1$) and $n=0$ we find from expressions (40), (41) and (55):

\begin{equation}
R_0(\alpha )\simeq \frac{2\cos ^2\alpha +\pi \beta _b^2\gamma _b^2\sin
^2\alpha }{\sqrt{\left( 1-\gamma _b^2\sin ^2\alpha \right) \left[ 1+\left(
\gamma _b^2-2\right) \sin ^2\alpha \right] }-2\beta _b^2\gamma _b^2\sin
^2\alpha \arcsin \left( \sqrt{\frac{1-\gamma _b^2\sin ^2\alpha }{2\cos
^2\alpha }}\right) },
\end{equation}
that for two limiting values of the angle $\alpha $ takes on the following
form:

\begin{equation}
R_0(\alpha )\simeq \left\{ 
\begin{array}{l}
2;\quad \alpha \rightarrow 0 \\ 
\frac{3(2+\pi )\left( 1-\gamma _b^{-2}\right) ^{3/2}}{2\sqrt{2}\left(
1-\gamma _b^2\sin ^2\alpha \right) ^{3/2}};\quad \alpha \rightarrow \Delta
\end{array}
\right. .
\end{equation}

The maximum wake field as well as the transformer ratio $R$ grow
monotonically with increasing density of bunch, the maximum value of the
field being achieved at $n_b/n_0=b_0$ and being determined by Eq. (61).

As was mentioned above, at $n_b/n_0>b_0$ the thickness of the bunch is
restricted by the value $d_{\max }$ (see Eq. (63)). If the density of bunch
is in the interval $b_0<n_b/n_0<b_c$, then one can obtain the maximum value
of decelerating field from expressions (40) and (41), where the expression $%
(d-\widetilde{\xi })/\lambda _p$ itself is determined by Eq. (25) with $\rho
=-\widetilde{\rho }$. For large densities of the bunch $n_b/n_0>b_c$ the
longitudinal field inside the bunch monotonically decrease from a definite
value at the first boundary to zero at the second boundary of the bunch.
Then the maximum value of decelerating field inside the bunch is equal to
the value of field at the first boundary (when $\xi =0$) and is determined
by expression (42). In the range of $n_b/n_0>b_0$ as a whole, the maximum
value of longitudinal wake field is $\left| {\cal E}_{1z}\right| $ (see Eq.
(55)), and, hence, in the last two cases (Eq. (42) and Eq. (55)) instead of $%
\rho _0$ and $d$ one is to put their maximum allowable values (see Eqs.
(63)-(65)).

In the limit of large densities of the bunch ($n_b/n_0\gg b_c$) the
following expression is obtained from Eq. (63):

\begin{equation}
d_{\max }\simeq \frac{\lambda _p}{\left( n_b/n_0\right) }\sqrt{\frac 2{\cos
^2\alpha }\sqrt{1-\gamma _b^2\sin ^2\alpha }\left( \gamma _b\cos \alpha -%
\sqrt{1-\gamma _b^2\sin ^2\alpha }\right) },
\end{equation}
that enables one to find asymptotic form of the transformer ratio for large
densities of the bunch:

\begin{equation}
R_\infty (\alpha )\simeq \frac{1+\left( \gamma _b^2-2\right) \sin ^2\alpha }{%
1-\gamma _b^2\sin ^2\alpha }.
\end{equation}

As follows from expressions (77), (78) and (80), in the presence of external
strong magnetic field in plasma, the transformer ratio increases and for $%
\alpha \lesssim \Delta $ much exceeds the value of $R$ at $B_0=0$.

In Figs. 8-10 the dependence of $R$ on the density of bunch ($n_b/n_0)$
(Figs. 8, 9) and on the angle $\alpha $ (Fig. 10) is shown for $\gamma _b=60$%
. In Fig. 8 the dotted, dashed and solid lines correspond to the following
values of the angle $\alpha $ respectively: $\alpha =0$, $\alpha =0.2\Delta $
and $\alpha =0.4\Delta $. In Fig. 9 the dotted and solid lines have been
plotted for $\alpha =0.6\Delta $ and $\alpha =0.8\Delta $ respectively. In
Fig. 10 the solid line corresponds to $n_b/n_0=0.7$, the dashed and dotted
lines respectively to $n_b/n_0=1$ and to $n_b/n_0=1.3$.

One can assure that the dotted line in Fig. 8, for which $\alpha =0$,
coincides with an analogous result obtained in Ref. [9] in the absence of
external magnetic field.

Thus, it follows from the above analysis and from Figs. 8-10 that the
presence of strong external magnetic field increases the amplitude of wake
waves excited by the one-dimensional bunch as well as the transformer ratio.
This effect is most strongly manifested in the case when the angle of
magnetic field orientation is close to $\Delta $ ($\alpha \lesssim \Delta $%
). Note also that the transformer ratio is maximum at $n_b/n_0\simeq b_0$,
and with increasing $\alpha $ the position of this value is shifted to lower
densities of the bunch. Hence, in the presence of strong magnetic field the
larger values of transformer ratio may be obtained at much lower densities
of the bunch than in the case of $B_0=0$.

\section{CONCLUSIONS}

In conclusion we shall give a brief summary of the results and discuss
possible restrictions on the validity of conclusions.

The behavior of one-dimensional electromagnetic waves excited in plasma in
the presence of external strong magnetic field, strongly depends on the
orientation of bunch motion with respect to the direction of magnetic field.
At large values of angle $\alpha $ ($\alpha \geq \Delta =\arcsin (1/\gamma
_b)$) the density and momentum of plasma electrons , as well as the excited
electromagnetic fields vary in the space monotonically. Here the
electromagnetic field increases linearly with the distance from the bunch.
Obviously, this result is valid only in the one-dimensional model and in
case of an allowance for the transverse size of the bunch (i.e., in the
three-dimensional case) the fields will tend to zero at large distances. In
Ref. [16] an excitation of three-dimensional linear wake waves was
considered in the magnetized plasma. It was shown here that at large
distances from the driving bunch the fields nevertheless tend to zero
although slowly (by the power-law dependence). So, the linear increase of
one-dimensional wake fields behind the bunch is limited by the distance that
is proportional to the cross-size of the bunch.

The allowance for the transverse size of the bunch results also in the
following additional restrictions [17]: $\gamma _b^2\lambda _\rho ^2\ll a^2$
and $r^2\ll a^2$, where $a$ is the transverse size of the bunch and $r$ is
the transverse coordinate.

At small values of the angle $\alpha $ ($\alpha <\Delta $) a periodic wake
fields is excited behind the bunch, the amplitude of which increases when $%
\alpha \rightarrow \Delta $. In this case the transformer ratio also
increases, the position of its maximum being displaced to lower densities of
the bunch.

It follows from the aforesaid that optimal conditions for the excitation of
wake waves are $d\lesssim \lambda _{{\rm in}}/2$ and $n_b/n_0\simeq
b_0(\alpha )$, where $b_0(\alpha )$ is determined by expression (60).

Another feature due to the presence of the magnetic field is the excitation
of transverse electric and magnetic fields, that are absent in the
one-dimensional case at $B_0=0$ [4-10].

The allowance for thermal motion of plasma electrons leads to inessential
corrections to wake wave characteristics at temperatures practically
achievable under laboratory conditions [18].

\vspace{1.0in}

\begin{center}
{\bf ACKNOWLEDGMENT}
\end{center}

The work was supported by the International Science and Technology Center
under Project No. A-013.

\newpage 

REFERENCES

\begin{enumerate}
\item  P. Chen, J. M. Dawson, R. W. Huff and T.Katsouleas, {\it Phys. Rev.
Lett.}, {\bf 54}, 693 (1985).

\item  Ya. B. Fainberg, {\it Proc. Symp. on Collect. Acc.}, CERN, {\bf 1},
84 (1956).

\item  J. B. Rosenzweig, D. B. Cline, B. Cole et al., {\it Bull. Am. Phys.
Soc.}, {\bf 32}, 1787 (1987).

\item  J. B. Rosenzweig, {\it Phys. Rev. Lett.}, {\bf 58}, 555 (1987).

\item  J. B. Rosenzweig, {\it IEEE Trans. Plasma Sci.}, {\bf 15}, 186 (1987).

\item  J. B. Rosenzweig, {\it Phys. Rev. A}, {\bf 38}, 3634 (1988).

\item  R. D. Ruth, A. Chao, P. L. Morton and P. B. Wilson, {\it Part. Acc.}, 
{\bf 17}, 171 (1985).

\item  A. Ts. Amatuni, E. V. Sekhpossian and S. S. Elbakian, {\it Fizika
Plazmy}, {\bf 12,} 1145 (1986).

\item  A. Ts. Amatuni, S. S. Elbakian, E. V. Sekhpossian and R. O. Abramian, 
{\it Part. Acc}., {\bf 41}, 153 (1993).

\item  A. G. Khachatryan, {\it Phys. Plasmas}, {\bf 4}, 4136 (1997).

\item  E. Esarey, P. Sprangle, J. Krall and A. Ting, {\it IEEE Trans. Plasma
Sci}., {\bf 24,} 252 (1996).

\item  Ya. B. Fainberg, {\it Plasma Phys. Rep.}, {\bf 23}, 251 (1997).

\item  J. M. Dawson, {\it Phys. Rev.}, {\bf 133}, 383 (1959).

\item  N. E. Andreev, L. M. Gorbunov and R. R. Ramazashvili, {\it Plasma
Phys. Rep.}, {\bf 23}, 277 (1997).

\item  A. Ts. Amatuni, S. S. Elbakian, E. M. Laziev et al., {\it Part. Acc}%
., {\bf 32}, 221 (1990).

\item  A. Ts. Amatuni, E. V. Sekhpossian, A. G. Khachatryan and S. S.
Elbakian, {\it Fizika Plazmy}, {\bf 21}, 1000 (1995).

\item  A. Ts. Amatuni, E. V. Sekhpossian and S. S. Elbakian, {\it Journal of
Contemporary Physics} (National Academy of Sciences of Armenia), {\bf 25}, 1
(1990), Allerton Press, Inc.

\item  A. Ts. Amatuni, S. S. Elbakian and E. V. Sekhpossian, {\it Part. Acc}%
., {\bf 36}, 241 (1992).

\newpage 
\end{enumerate}

\begin{center}
{\bf FIGURE\ CAPTIONS}
\end{center}

FIGURE 1: The dependence of the $\lambda _{{\rm in}}/\lambda _p$ on $n_b/n_0$
($0\leq n_b/n_0\leq b_1$) for $\gamma _b=50$, $\alpha =0$ (solid line), $%
\alpha =0.2\Delta $ (dashed line) and $\alpha =0.4\Delta $ (dotted line).

FIGURE 2: The dependence of the $\lambda _{{\rm out}}/\lambda _p$ on $\rho
_{\max }$ for $\gamma _b=50$, $\alpha =0$ (solid line), $\alpha =0.6\Delta $
(dashed line) and $\alpha =0.8\Delta $ (dotted line).

FIGURE 3: The dependence of the function $eE_z(\xi )/mc\omega _p$ on $\xi
/\lambda _p$ for $n_b/n_0=b_0(\alpha )$, $d=\lambda _{{\rm in}}/2$, $\gamma
_b=60$, $\alpha =0$ (solid line), $\alpha =0.2\Delta $ (dashed line) and $%
\alpha =0.3\Delta $ (dotted line).

FIGURE 4: The dependence of the function $eE_x(\xi )/mc\omega _p$ on $\xi
/\lambda _p$ for $n_b/n_0=b_0(\alpha )$, $d=\lambda _{{\rm in}}/2$, $\gamma
_b=60$, $\alpha =0.1\Delta $, $\alpha =0.15\Delta $ and $\alpha =0.2\Delta $.

FIGURE 5: The dependence of the function $eE_z(\xi )/mc\omega _p$ on $\xi
/\lambda _p$ for $n_b/n_0=0.8b_0(\alpha )$, $d=\lambda _{{\rm in}}/2$, $%
\gamma _b=60$, $\alpha =0$ (solid line), $\alpha =0.2\Delta $ (dashed line)
and $\alpha =0.4\Delta $ (dotted line).

FIGURE 6: The dependence of the function $eE_x(\xi )/mc\omega _p$ on $\xi
/\lambda _p$ for $n_b/n_0=0.8b_0(\alpha )$, $d=\lambda _{{\rm in}}/2$, $%
\gamma _b=60$, $\alpha =0.1\Delta $ (solid line), $\alpha =0.15\Delta $
(dashed line) and $\alpha =0.2\Delta $ (dotted line).

FIGURE 7: The dependence of the $d_{\max }/\lambda _p$ on $n_b/n_0$ ($%
n_b/n_0\geq b_0$) for $\gamma _b=60$, $\alpha =0$ (solid line), $\alpha
=0.4\Delta $ (dashed line) and $\alpha =0.8\Delta $ (dotted line).

FIGURE 8: The dependence of the transformer ratio $R$ on $n_b/n_0$ for $%
\gamma _b=60$, $\alpha =0$ (solid line), $\alpha =0.2\Delta $ (dashed line)
and $\alpha =0.4\Delta $ (dotted line).

FIGURE 9: The dependence of the transformer ratio $R$ on $n_b/n_0$ for $%
\gamma _b=60$, $\alpha =0.6\Delta $ (dotted line) and $\alpha =0.8\Delta $
(solid line).

FIGURE 10: The dependence of the transformer ratio $R$ on anisotropy angle $%
\alpha $ for $\gamma _b=60$, $n_b/n_0=0.7$ (solid line), $n_b/n_0=1$ (dashed
line) and $n_b/n_0=1.3$ (dotted line).

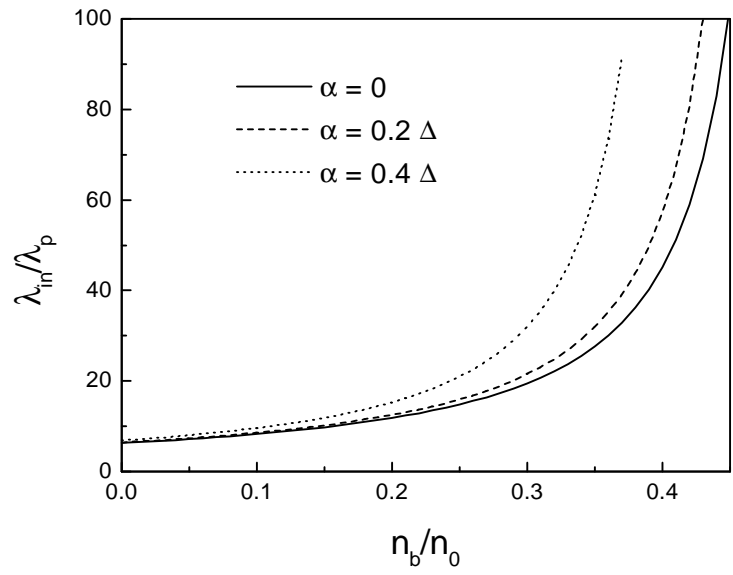
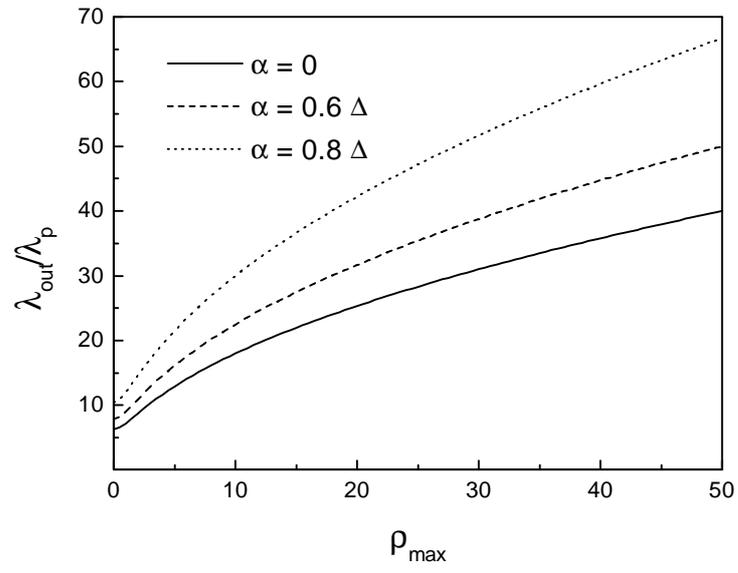
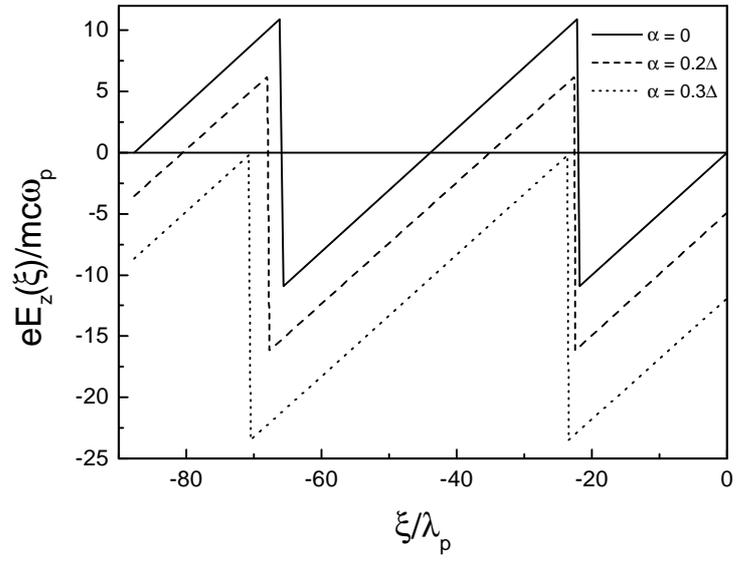
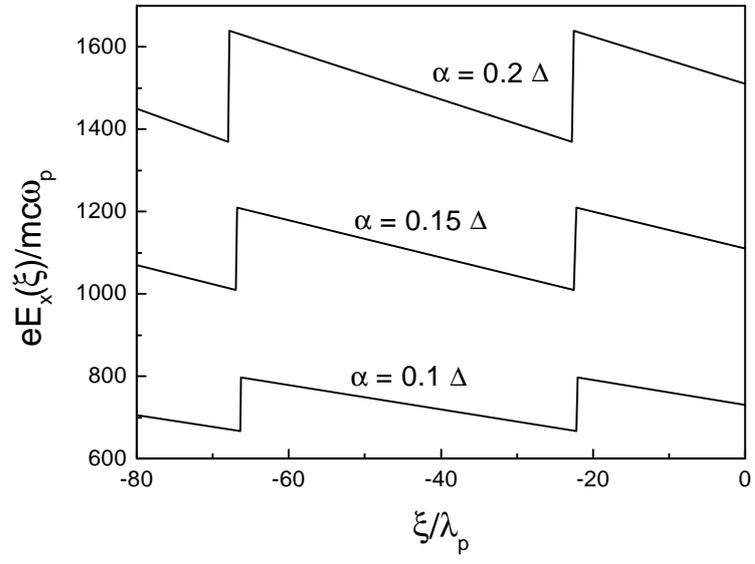
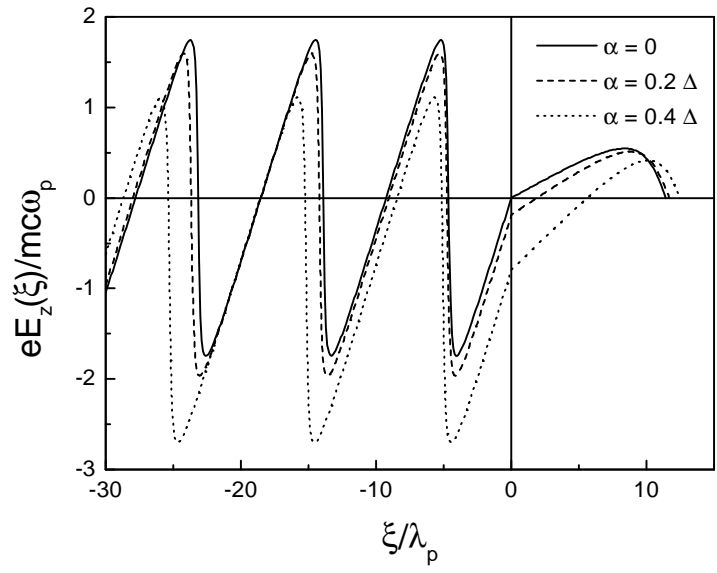
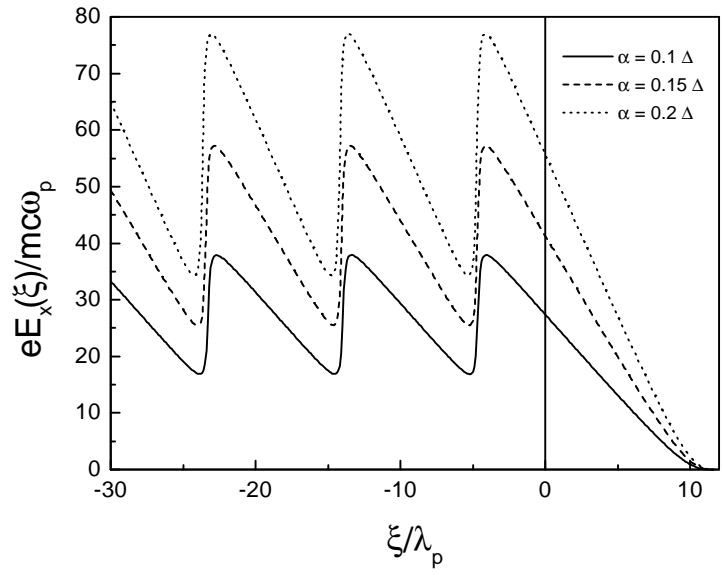
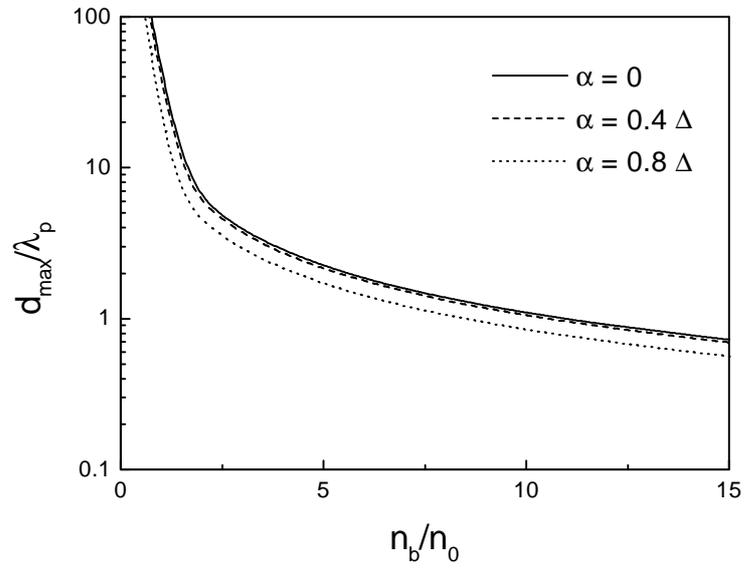
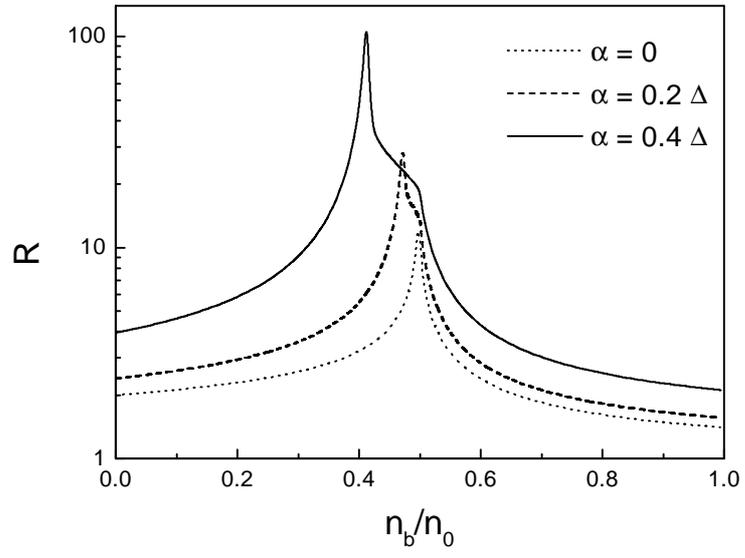
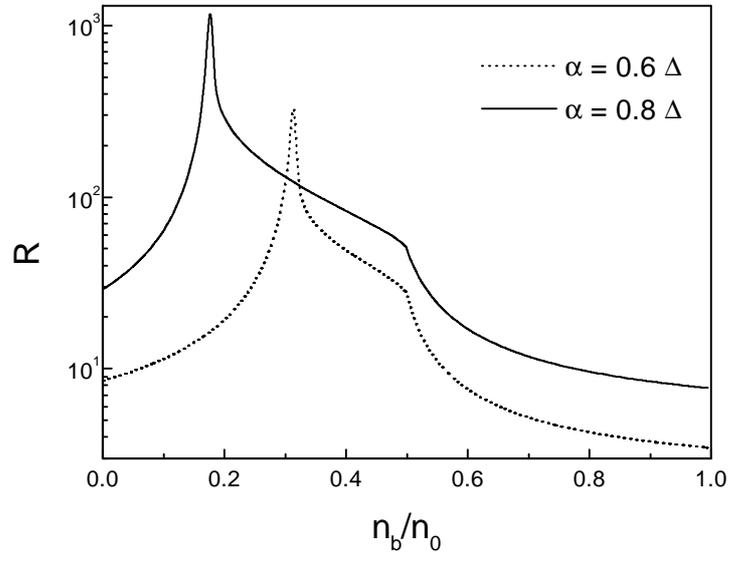
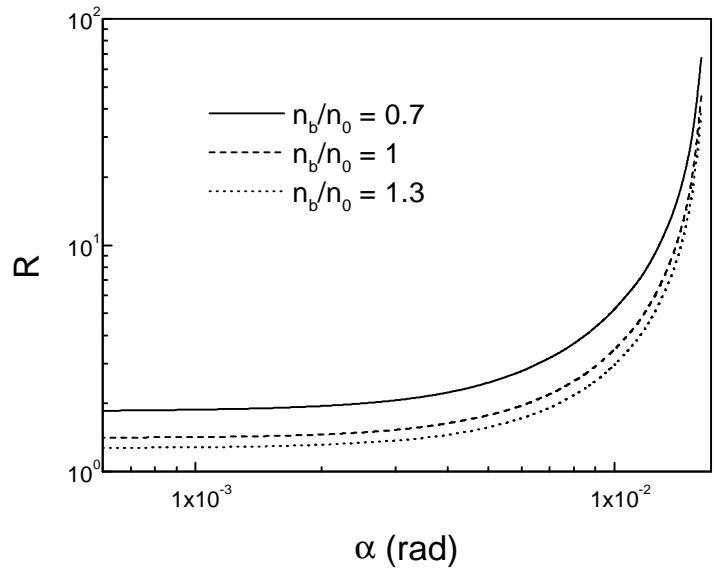
\end{document}